\documentstyle[11pt]{article}
\def\slashchar#1{\setbox0=\hbox{$#1$}           % set a box for #1
   \dimen0=\wd0                                 % and get its size
   \setbox1=\hbox{/} \dimen1=\wd1               % get size of /
   \ifdim\dimen0>\dimen1                        % #1 is bigger
      \rlap{\hbox to \dimen0{\hfil/\hfil}}      % so center / in box
      #1                                        % and print #1
   \else                                        % / is bigger
      \rlap{\hbox to \dimen1{\hfil$#1$\hfil}}   % so center #1
      /                                         % and print /
   \fi}
\begin{document}
\begin{titlepage}
\begin{center}
July, 1998      \hfill     BUHEP-98-20\\
\vskip 0.2 in
{\Large \bf Compactification for a Three-Brane Universe}

\vskip .2 in
         {\bf   Raman Sundrum\footnote{email: sundrum@budoe.bu.edu.}}
        \vskip 0.3 cm
       {\it Department of Physics \\
Boston University \\
Boston, MA 02215, USA}
 \vskip 0.7 cm 
%\vskip 0.3 in

\begin{abstract}
A  fully realistic and systematic effective field theory model 
of a 3-brane universe is constructed. It consists of a 
six-dimensional gravitating spacetime, containing several, 
approximately parallel (3+1)-dimensional defects, or ``3-branes''. 
The Standard Model particles are confined to live on one of the 
3-branes while different four-dimensional field theories may 
inhabit the others, in literally a case of ``parallel universes''. 
The effective field theory is valid up to the six-dimensional 
Planck scale, where it must be replaced by a more fundamental 
theory of gravity and 3-brane structure. Each 3-brane induces a 
conical geometry in the two dimensions transverse to it.
Collectively, the curvature induced by the 3-branes can compactify  
the extra dimensions into a space of spherical topology. It is 
possible to take the six-dimensional Planck scale to be not much 
larger than the weak scale, and the compact space not much smaller 
than a millimeter, thereby realizing the recent proposal by 
Arkani-Hamed, Dimopoulos and Dvali for eliminating the Gauge 
Hierearchy Problem. In this case, an extra force is required to 
stabilize the compact space against collapse. This is provided by 
a six-dimensional (compact) U(1) gauge field with a magnetic flux 
quantum trapped in the compact space. The nature of the Cosmological 
Constant Problem in this scenario is discussed. 
\end{abstract}
\end{center}

\end{titlepage}

\section{Introduction}

It is usually assumed that the fundamental dynamical 
scale underlying gravity is
the Planck scale, 
\begin{equation}
M_{Pl} \sim 10^{18} {\rm GeV},
\end{equation}
set by the observed value of Newton's constant. If so, one is faced with
the problem of understanding the mechanism which stabilizes the very
large hierarchy between this scale and the electroweak scale, $v = 246$
GeV. Recently however, it has been proposed \cite{stanford1} that the
dynamical scale of gravity, $M$, is not much larger than the weak scale,
thereby eliminating the usual hierarchy problem! This is accomplished by
taking general relativity to be fundamentally {\it six}-dimensional,
with two large compact dimensions, and identifying $M$ with the
six-dimensional Planck mass. Using the standard relation
\cite{stanford1} (also see Section 6 within),
\begin{equation}
M_{Pl}^2 \sim {\cal A} M^4, 
\end{equation}
where ${\cal A}$ is the area of the compact two-dimensional space,
one finds that if $M$ is not much larger
than the weak scale, then the typical (linear) dimension of the compact
space is not much smaller than a millimeter. That is, the
compactification mass scale is not much larger than $10^{-4}$ eV.
Ref. \cite{stanford1} proposed that the reason we do not experimentally 
observe finely spaced Kaluza-Klein excitations of the Standard Model
(SM) particles is because the entire SM is confined to a
$(3+1)$-dimensional defect, which we will refer to as a ``3-brane'',
 which is point-like in the two-dimensional
compact space and extended in the non-compact directions. On the other
hand, gravity is not confined in this manner and light Kaluza-Klein
excitations of the graviton are present. The net result is that gravity
 is effectively described by four-dimensional general relativity at
 distances larger than a millimeter, but at shorter distances the
 Kaluza-Klein excitations propagate and gravity reveals its
 six-dimensional nature. To date, gravity has only been tested down to
 a distance of a centimeter with no sign of extra dimensions,
 but if we are lucky the six-dimensional 
 transition may appear in  upcoming sub-millimeter tests of gravity
 \cite{stanford1} \cite{price}.

There is an ongoing effort to theoretically realize a 
phenomenologically acceptable version of the above scenario,
either within quantum field theory or within string 
theory \cite{stanford1} \cite{dvali} \cite{stanford2} \cite{tye}. 
Related ideas involving 3-brane universes and/or
 relatively low compactification mass scales
appear in refs. [6 -- 15]. The purpose of the present paper is to
 construct a realistic model of a 3-brane universe using the 
effective field theory methods developed in ref. \cite{me}, focussing on
the compactification mechanism.  This
approach is analogous to the chiral lagrangian approach to the
soft pion sector of the strong interactions. Just as the chiral
lagrangian describes the most general structure of the 
low-energy interactions among (pseudo)-Nambu-Goldstone bosons, without
explicitly describing the mechanism that gave rise to the
associated spontaneous symmetry-breaking, the effective field theory we
use here will describe the general structure of low-energy interactions
among the 3-brane fluctuations, six-dimensional gravity and the SM
fields, without explictly describing the mechanism that gave rise to the
3-brane and the fields living on it. Just as the chiral lagrangian
description is valid up to energies at which the detailed QCD mechanism
for chiral symmetry-breaking becomes important, the effective theory we
will use is valid up to energies of order $M$, at which point the
internal structure of the 3-brane and the physics of strongly-coupled
gravity become important. 

The basic model proposed in this paper consists of a six-dimensional
gravitating spacetime, containing several, approximately parallel
3-branes. They act as sources for the curvature
needed to compactify the extra dimensions into a space of spherical
topology. Each of the 3-branes may be inhabited by a separate
four-dimensional quantum field theory, one of which is the familiar
SM. These  parallel, four-dimensional sub-universes 
interact weakly with each other via the bulk six-dimensional gravity, so
that they can be considered as hidden sectors relative to each other. This is
qualitatively similar to the ideas put forth in refs. \cite{horava}. 

This paper is organized so as to progressively build up to a fully
realistic model. 
Section 2 briefly reviews the necessary
effective field theory formalism detailed in ref. \cite{me}. Section 3
describes the effective field theory that results
from integrating out the massive SM physics. Section 4
deals with the case of a single 3-brane in six fully non-compact
dimensions. The classical equations of motion reveal that the 
3-brane induces a conical geometry in the two transverse
dimensions. In Section 5, the cones from several 3-branes are patched
together to fully compactify the extra dimensions. In Section 6, the
effective field theory below the compactification mass scale is derived.
It is pointed out that when the vev of the 
effective field corresponding to the size of
the compact space is large, this field 
can mediate effects in conflict with experimental
post-Newtonian gravitational tests. This problem is resolved in Section 7
by introducing a six-dimensional abelian gauge field with a non-trivial
magnetic flux through the compact space. Section 8 discusses
the nature of the Cosmological Constant Problem in the present
scenario. Section 9 contains the final discussion.

\section{The Standard Model on a 3-brane}

This section summarizes some of the key aspects of the 3-brane effective field
theory formalism described in ref. \cite{me}. 

Our starting point will be the action governing the SM fields on a
3-brane, which in turn is coupled to six-dimensional ``bulk'' gravity 
\cite{me}, 
\begin{eqnarray}
S &=& S_{3-brane} ~+~ S_{bulk}, \\
S_{3-brane} &=& \int d^4 x~ \sqrt{-g} ~\{-f_0^4 ~-~ g^{\mu \nu} D_{\mu}
H^* D_{\nu} H ~-~ V(H, H^*) \nonumber \\
-~ \frac{1}{4}~ g^{\mu \rho}
g^{\nu \sigma}~ F_{\mu \nu} F_{\rho \sigma} 
&+& \overline{\psi}_L i e^{\mu}_{~\alpha}~ \sigma^{\alpha} D_{\mu}
\psi_L ~+~ y H \psi_L \psi_L ~+~ {\rm h.c.} ~+~ ...\}, \\ 
S_{bulk} &=& \int d^6 X~ \sqrt{-G} ~\{- \Lambda_0 ~-~ 2 M_0^4 R
\nonumber \\
&-& \frac{1}{4}~ {\cal F}^{MN} {\cal F}_{MN} ~+~ ...\}.
\end{eqnarray}
The SM scalar, chiral spinor and vector fields are denoted $H,
\psi_L, A_{\mu}$ respectively, the last of these being used to form the
covariant derivatives and the gauge field strength, $F_{\mu \nu}$. 
(Gauge and flavor indices have not been explicitly written.) The SM fields are
functions of 
intrinsic coordinates on the
3-brane, $x^{\mu ~=~ 0,...,3}$. The gravitational field is the
six-dimensional metric, $G_{MN}$, used to construct the six-dimensional
curvature scalar, $R$, and is a function of coordinates for
the bulk spacetime, $X^{M ~=~ 0,...,5}$. A six-dimensional compact
$U(1)$  gauge
field is also included, with field strength ${\cal F}_{MN}(X)$. It will
not play an important role until Section 7. 
The 3-brane embedding in the bulk
spacetime is given by fields, $Y^M(x)$. The SM fields ``feel'' a
four-dimensional metric on the 3-brane induced by this embedding, given
by,
\begin{equation}
g_{\mu \nu}(x) = G_{MN}(Y(x)) ~\partial_{\mu}Y^{M} ~\partial_{\nu} Y^N.
\end{equation}
(The case of chiral fermions is more subtle, involving an induced vierbein,
$e^{\alpha}_{~\mu}$. It is given careful treatment in ref. \cite{me},
but we will not need the details here.) The dimensionful constants, 
$f_0^4$, $\Lambda_0$ and $M_0$, are the ``bare'' 3-brane tension,
six-dimensional cosmological constant and six-dimensional Planck mass
respectively. In this paper we will consider the case where $f_0 \sim
{\cal O}(M_0)$.

The terms in eqs. (4) and (5) are
the lowest dimension operators which are invariant under both general
$X$-coordinate transformations and $x$-coordinate transformations, the
ellipses containing higher-dimension invariants suppressed by powers of
$M_0$. The resulting theory, written in terms of canonical
fields, is necessarily non-renormalizable
and must be treated by the methods of effective field theory, the
effective theory being valid up to energies of order $M_0$.
 This scale constrains both the allowed
 energy-momenta in physical processes, and also the size of metric fluctuations
away from six-dimensional Minkowski space and 3-brane fluctuations away
from a flat four-dimensional hypersurface.
Physics at higher energies can only be
 understood within a more fundamental theory, describing the internal
 structure of the 3-brane and strongly-coupled  gravity. The fact that SM
 experiments are not sensitive to such exotic physics indicates that 
$M_0$ (and hence $f_0$) are at least larger than the weak scale (appearing
 in the SM potential, $V(H, H^*)$).

Finally, consider the embedding fields, $Y^M(x)$. 
Because of the coordinate invariances, not all of the $Y^M$ are 
physical. A convenient gauge-fixing (in the effective theory's domain of
validity) is provided by choosing,
\begin{eqnarray}
Y^{\mu}(x) &=& x^{\mu}, ~~Y^{m = 4,5}(x) ~{\rm arbitrary}.
\end{eqnarray}
The two physical fields, $Y^m$, acquire explicit kinetic terms and interactions
upon expanding eq. (4) for small fluctuations, using eq. (6). They
 appear derivatively coupled because they are the Nambu-Goldstone
modes corresponding to spontaneous breaking of transverse translations
by the 3-brane. We will generally use lower-case Roman letters, $m, n,
... = 4, 5$, to denote these transverse directions.

\section{Infrared Dynamics of Gravity and the 3-brane}

Let us now imagine integrating out the 
physics of the effective theory described above, down to the far
infrared. In particular, all the massive SM particles are completely
integrated out. We are left with an effective theory valid at very low
energies, consisting of six-dimensional
gravity and the abelian gauge field, 
the 3-brane embedding fields, and massless SM particles. 
Now at these energies 
the massless SM particles are essentially
decoupled from each other and from the gravitational and
(derivatively-coupled) embedding fields. Therefore, since we are
more interested  in the dynamics of the 3-brane itself, 
we can drop reference to the massless SM fields since they have a
negligible effect. Similarly we can ignore the six-dimensional gauge
field (which has no sources in this paper). 
Then the general form of this infrared effective
theory is given by eq. (3), where now, 
\begin{eqnarray}
S_{3-brane} &=& \int d^4 x~ \sqrt{-g} ~\{-f^4 ~+~ ...\}, \\ 
S_{bulk} &=& \int d^6 X~ \sqrt{-G} ~\{- \Lambda ~-~ 2 M^4 R ~+
...\}.
\end{eqnarray}
The higher dimension terms in the ellipses are irrelevant in
the far infrared.
The dimensionful constants, $f^4$, $\Lambda$ and $M$ are the bottom-line
renormalized 3-brane tension, six-dimensional cosmological constant
and six-dimensional Planck constant respectively. The fact that we took
$f_0 \sim {\cal O}(M_0)$ implies that, barring fine cancellations, $f
\sim {\cal O}(M)$.   For now
we shall assume that $\Lambda = 0$, but will take it to be non-zero but
small in Sections 6 and 7. 

\section{A Flat 3-brane Solution}

The dynamics of the 3-brane and six-dimensional gravity is weakly
coupled in the far infrared, and well-approximated by the
classical equations of motion. We will look for a static 
solution of the following form, 
\begin{eqnarray}
Y^m &=& \overline{Y}^m = {\rm constant}, ~~Y^{\mu}(x) = x^{\mu}, \nonumber
\\
ds^2 &\equiv& G_{MN} ~dX^M~ dX^N \nonumber \\
&=&  \eta_{\mu \nu}~ dx^{\mu}~dx^{\nu} + {\cal G}_{mn}(X^{4}, X^5)~
dX^m~ dX^n.
\end{eqnarray}
That is, the bulk spacetime has the form, ${\rm Mink}_4 \times {\cal M}_2$,
where ${\rm Mink}_4$ is  four-dimensional Minkowski
space and ${\cal M}_2$ is a two-dimensional manifold. In this section we
will take ${\cal M}_2$ to have non-compact planar topology. The
3-brane is embedded along the Minkowski directions and at
some point, $\overline{Y}^m$, in ${\cal M}_2$. Note that, by eq. (6), 
with such an embedding the SM fields would feel an induced Minkowski metric
$g_{\mu \nu} =  \eta_{\mu \nu}$. Also notice that our ansatz for $Y$
satisfies the gauge condition, eq. (7).

When obtaining the classical equations of motion by functionally
differentiating the action,
it is legitimate to derive the $Y^m$ equations of motion by first setting
$G_{MN}$ to its ansatz form, 
and then to derive the metric equations of
motion by first setting $Y$ to its ansatz form. 
The $Y^m$ equations of
motion are then, 
\begin{equation}
\partial_{\mu} [\sqrt{-g} ~g^{\mu \nu}~ {\cal G}_{mn}(Y(x))~
\partial_{\nu} Y^n] = 0, 
\end{equation}
where, 
\begin{equation}
g_{\mu \nu} = \eta_{\mu \nu} ~+ ~{\cal G}_{mn}(Y)~ \partial_{\mu} Y^m~
\partial_{\nu} Y^n.
\end{equation}
Clearly, $Y^n = \overline{Y}^n =$ constant provides a solution to these 
equations.

The metric equations of motion
(six-dimensional Einstein equations) are,
\begin{equation}
\sqrt{-G} (R_{MN} ~-~ \frac{1}{2} R~ G_{MN})(X) = \frac{f^4}{4 M^4}
~G_{M \mu}(X)~ \eta^{\mu \nu} ~G_{\nu N}(X) ~\delta^2(X^m -
\overline{Y}^m).
\end{equation}
Now let us try the metric ansatz. The only non-trivial components of the
curvature tensor can be $R_{mn}$, and Einstein's equations split into
two, 
\begin{eqnarray}
\sqrt{\cal G} R^{(2)} &=& - \frac{f^4}{2 M^4} \delta^2(X^m -
\overline{Y}^m)  \\
R_{mn} &-& \frac{1}{2} R^{(2)}~ {\cal G}_{mn} = 0,
\end{eqnarray}
where $R^{(2)}$ denotes the two-dimensional curvature scalar constructed
from ${\cal G}_{mn}$. Eq. (15) holds identically for any two-dimensional
metric, ${\cal G}_{mn}$. Eq. (14) is closely analogous to 
Einstein's equations in
$(2+1)$ dimensions in the presence of a static particle source \cite{deser},
and  has a very simple solution: ${\cal G}_{mn}$
corresponds to a conical geometry on ${\cal M}_2$, with the tip of the
cone at $\overline{Y}^m$. The deficit angle of the cone is given by 
\begin{eqnarray}
\delta =  \frac{f^4}{4 M^4}.
\end{eqnarray}
Although we will not need it here, an explicit form for this metric can be
given in radial coordinates centered at $\overline{Y}^m$, that is $X^4 -
\overline{Y}^4 \equiv \rho, ~X^5 - \overline{Y}^5 \equiv \phi$:
\begin{equation}
{\cal G}_{\rho \rho} = 1, ~ {\cal G}_{\rho \phi} = {\cal G}_{\phi
  \rho} = 0, ~ {\cal G}_{\phi \phi} = (1 - \frac{f^4}{8 \pi M^4})^2~
\rho^2.
\end{equation}
Away from the 3-brane, the bulk spacetime has Minkowskian geometry.

\section{Compactification with Several 3-branes}

Of course, the above solution does not provide a realistic background
spacetime, because gravitational fluctuations can propagate freely in the six
non-compact dimensions. For example, this leads to a $1/r^4$
Newtonian force instead of the experimental $1/r^2$ law
\cite{stanford1}. To cure
this problem we will consider ${\cal M}_2$ to be compact, with spherical
topology. However, we must reconcile this with the fact that the static
3-brane considered in the previous section yields a locally flat ${\cal
  M}_2$ with a conical singularity at the 3-brane location. The simplest
way to proceed is to consider the case of several 3-branes, labelled by
an index $j$. Each of these 3-branes may be inhabited by different
four-dimensional field theories, one of which is the SM. At low enough
energies the details of these field theories are irrelevant, the 3-branes
are charactererized just by their renormalized tensions, $f^4_j$. 
We will look
for a solution to the classical equations of motion using the same
ansatz for the metric as in the previous section and with each 3-brane
again extended in the Minkowski directions and occupying a fixed point,
$\overline{Y}^m_j$, in ${\cal M}_2$. This configuration is a case of ``parallel
universes'' linked only by the higher-dimensional gravity.

As in the previous section, it is straightforward to see that the ansatz
satisfies the $Y_j$ equations of motion. The non-trivial Einstein
equation generalizes
to,
\begin{eqnarray}
\sqrt{\cal G} R^{(2)} &=& - \sum_j ~\frac{f^4_j}{2 M^4} \delta^2(X^m -
\overline{Y}^m_j).
\end{eqnarray}
The solution is now analogous to the case of several static point masses
in $(2+1)$-dimensional gravity on a space of spherical topology
\cite{deser}. The geometry is flat everywhere in ${\cal M}_2$ except at the
locations of the 3-branes, where there are conical singularities with
deficit angles, 
\begin{equation}
\delta_j = \frac{f^4_j}{4 M^4}.
\end{equation}

However this static solution is not generally possible because of the
constraint provided by the Gauss-Bonnet Theorem for spherical topology, 
\begin{equation}
\int_{{\cal M}_2} d X^4~ d X^5~ \sqrt{\cal G} R^{(2)} = - 8 \pi.
\end{equation}
By eq. (18) this implies that the static solution is only possible if the
3-brane tensions satisfy the sum rule, 
\begin{equation}
\sum_j~ \frac{f^4_j}{4 M^4} = 4 \pi.
\end{equation}
That is, according to eq. (18), 
 the conical singularities are the only source of curvature for
${\cal M}_2$, and the deficit angles must add up to
$4 \pi$ in order to yield a surface of spherical topology. 
A simple example of such an ${\cal M}_2$ is the surface of a
tetrahedron, where the four vertices correspond to the positions of four
3-branes. 

As will be seen in Section 6 and discussed further in Section 8, 
the fine-tuning of 3-brane tensions
to satisfy eq. (21)  is just the re-incarnation of the
Cosmological Constant Problem in the present context. Granting this
fine-tuning, we have
a  minimal mechanism for compactification of the
higher dimensions,  namely the curvature induced by the 3-branes
themselves.

\section{Effective Field Theory below the Compactification Scale}

Let us now derive the effective field theory for the
massless degrees of freedom after compactification in the manner
described in the previous section. As usual when higher dimensions are
compactified, the higher dimensional fields, in the present case just
the six-dimensional metric (neglecting the abelian gauge field for now),
give rise to a Kaluza-Klein tower of
four-dimensional states, most of which acquire masses of order the
compactification mass scale. The massless states correspond to
``zero-modes'' of the compactified configuration. The 3-brane is an added
source of massless fields, namely the $Y^m$ fields 
and the massless SM fields.
We will neglect the massless SM fields 
here since their inclusion is rather trivial. Obviously
the fields on the 3-brane do not give rise to any massive Kaluza-Klein
states.

Let us first consider the six-dimensional metric tensor, whose
components can be
decomposed in four-dimensional Minkowski space as a tensor $G_{\mu
  \nu}(X)$, vector fields, $G_{\mu m}(X)$, and scalars,
$G_{mn}(X)$.  We found a global Minkowski
four-dimensional spacetime factor as part of our classical solution in
Section 5. We can deform this
continuously so that this spacetime factor is a curved manifold
described by a four-dimensional metric, 
\begin{equation}
G_{\mu \nu} = \overline{g}_{\mu \nu}(x), 
\end{equation}
in coordinates where $X^{\mu} \equiv x^{\mu}$. 
Non-trivial $X^{4,~5}$-dependence in $G_{\mu \nu}$ corresponds to
Kaluza-Klein excitations with masses of order the compactification scale.
 In
Kaluza-Klein theory, massless components of the $G_{\mu
  m}$ correspond to continuous isometries of the compactified space.  
In the present case however, there are 
no continuous isometries (for example, consider the case
where ${\cal M}_2$ is a tetrahedron). Therefore there are no massless 
$G_{\mu m}$ states, and below the compactification scale we
effectively have,
\begin{equation}
G_{\mu m} = 0.
\end{equation}

The fate of the $G_{mn}(X)$ is tied up with the 3-brane fields,
$Y^m(x)$. The identification of zero-modes is made somewhat ambiguous by
general coordinate invariance. In one convenient choice of  language, 
 we can observe that the metric on ${\cal M}_2$, given by $G_{mn} = {\cal
   G}_{mn}$, is the euclidean metric plus a number of conical
 singularities at the 3-brane positions, 
the deficit angles fixed by the 3-brane tensions
 according to eq. (19). Therefore the zero-modes are the 
 3-brane separations, $|Y_j - Y_k|$, measured with the euclidean metric,
 which then determine ${\cal G}_{mn}$. Note that by
eqs. (6, ~7), the induced metric on the $j$-th 3-brane is given by, 
\begin{equation}
g_{\mu \nu}(x) = \overline{g}_{\mu \nu}(x) ~+~ 
{\cal G}_{mn}(Y^4_j(x), Y^5_j(x)) ~\partial_{\mu} Y^m~
\partial_{\nu} Y^n. 
\end{equation}
This formula requires careful interpretation because ${\cal G}$ is
being evaluated at a position where it has a conical singularity. We
do not expect a 3-brane to ``feel'' the curvature singularity for which
it is itself the source, anymore than we expect this for a point
particle. Indeed, given  any physical 
ultraviolet regularization of our original
effective theory at scale $M$, such curvature singularities would be
smoothed out over distances of order $1/M$. In the 3-brane 
effective lagrangian, we are  using ${\cal G}_{mn}$ to measure
distances involved in 3-brane fluctuations. In the effective theory's
domain of validity the typical distances are much larger than $1/M$, so
the curvature singularity is unimportant. In the present situation,
throwing out the curvature singularity leaves us with,
\begin{equation}
g_{\mu \nu}(x) = \overline{g}_{\mu \nu}(x) ~+~ 
\delta_{mn} ~\partial_{\mu} Y^m_j~
\partial_{\nu} Y^n_j, 
\end{equation}
where $\delta_{mn}$ denotes the two-dimensional euclidean metric. 

The effective four-dimensional 
theory below the compactification scale is then given by substituting the
massless metric components into eqs. (8, ~9),  
\begin{eqnarray}
S &=& -~ \int d^4 x~ \sqrt{- \overline{g}} ~\{~ 
\sum _j [f^4_j ~+~  \frac{f^4_j}{2}
\delta_{mn}~ \overline{g}^{\mu \nu}~ \partial_{\mu} Y^m_j ~\partial_{\nu}
Y^n_j] \nonumber \\
&+& \int d X^4~ d X^5 \sqrt{\cal
  G} [2 M^4 R^{(4)} ~+~ 2 M^4 R^{(2)}] + ... ~\},
\end{eqnarray}
where $R^{(4)}$ is the four-dimensional curvature scalar due to
$\overline{g}_{\mu \nu}$, and $R^{(2)}$ is the two-dimensional 
curvature scalar due to ${\cal G}_{mn}$. The kinetic
terms for the $Y_j$ arose by expanding $\sqrt{-g}$ in eq. (8) about
$\overline{g}_{\mu \nu}$, using eq. (25). By the Gauss-Bonnet theorem,
eq. (20), and the fact that $R^{(4)}$ constructed from $\overline{g}_{\mu
  \nu}$ is independent of $X^{4, ~5}$, we get,
\begin{eqnarray}
S &=& - \int d^4 x~ \sqrt{- \overline{g}} \{ \sum_j f^4_j - 16 \pi M^4
~+~ \sum _j \frac{f^4_j}{2}
\delta_{mn}~ \overline{g}^{\mu \nu} \partial_{\mu} Y^m_j ~\partial_{\nu}
Y^n_j \nonumber \\
 &+& 2~ {\cal A}(Y) M^4 R^{(4)} + ...\},
\end{eqnarray}
where,
\begin{equation}
{\cal A}(Y) \equiv \int d X^4~ d X^5 \sqrt{\cal  G}, 
\end{equation}
is the area of ${\cal M}_2$ determined by the 3-brane 
separations. It is now clear that our previous sum rule requirement on
the 3-brane deficit angles, eq. (21), is precisely the tuning of
parameters necessary to cancel the effective 
four-dimensional cosmological constant. 

From eq. (27) we see that the effective four-dimensional Planck
constant is given by, 
\begin{equation}
M_{Pl}^2 = {\cal A}(Y) M^4.
\end{equation}
The fact that $M_{Pl}$ depends on massless fields 
leads to a
conflict with post-Newtonian experimental tests of general
relativity. See ref. \cite{damour} for a review. These tests are
sensitive to the tensorial nature of the macroscopic gravitational
force, and imply that the scalar admixture of the gravitational force
can be
at most a fraction of a percent. At first sight this does not appear to
pose a problem for us since the derivative couplings of the (canonically
normalized \cite{me}) $Y$ scalars to SM
states (our standard probes of gravity), 
as given by eqs. (4) and (6), are negligible at distances larger
than a centimeter where gravity is tested.\footnote{Recall that we are
  considering the case where  $f_j \sim {\cal O}(M)$ are
  larger than the weak scale. (That is, by eqs. (18) and (20), we are
  considering the case of several order one deficit angles adding up to
  $4 \pi$.) The scales $f_j$ suppress the derivative couplings of the
  canonically normalized $Y$ scalars.} However the problem is that a
$Y$-dependent four-dimensional Planck mass in eq. (27) corresponds to an
order one mixing of the (normalized) $Y$ scalars with the metric tensor,
$\overline{g}_{\mu \nu}$, which  leads to an unacceptable scalar admixture to
gravity of order one.\footnote{One can also  perform
 a Weyl transformation to eliminate
the field dependence from the Einstein action.  It then resurfaces in
direct gravitational strength couplings of the scalars to the SM.} 

To escape from this phenomenological 
problem we require a potential energy term for the compactified
area, ${\cal A}(Y)$, which stabilizes it and gives the corresponding
combination of $Y$-scalars a finite Yukawa range below a
centimeter. The simplest way to introduce a potential term is to
begin with a small positive six-dimensional cosmological constant,
$\Lambda$. At
the level of the effective theory below the compactification scale this
clearly leads to an extra term, 
\begin{eqnarray}
\delta S_{\Lambda}
 &=& - \int d^4 x~ \sqrt{- \overline{g}} ~{\cal A}(Y) \Lambda.
\end{eqnarray}
This term favors the reduction of ${\cal A}$. We now require a force
that opposes this reduction in order to obtain stability. 
It was proposed in refs. \cite{candelas}, that this can
be provided by intrinsically quantum mechanical matching corrections at the
compactification scale. Although this is not the method of
stabilization preferred in this paper, it is useful to briefly 
consider it first in order to understand its merits and problems.

The important point is that there is a tower of
Kaluza-Klein states of the graviton with masses set by the
compactification scale. Quantum loops of these states will then contribute to
the effective potential. Dimensional analysis suggest the rough form, 
\begin{equation}
\delta S_{quantum} = - 
\int d^4 x~ \sqrt{ - \overline{g}} ~\frac{k}{{\cal A}^2}. 
\end{equation}
 Indeed these are just the types of
corrections, with $k > 0$,  that are induced in more standard Kaluza-Klein
compactifications. For example, see ref. \cite{myers}.
We see that combining our original effective action with 
$\delta S_{\Lambda}$ and $\delta S_{quantum}$
gives a stabilizing effective potential for ${\cal A}(Y)$, 
\begin{equation}
{\cal V}_{eff} = {\cal A}(Y) \Lambda ~+~ \frac{k}{{\cal A}(Y)^2} ~+~
\sum_j f^4_j ~-~ 16 \pi M^4.
\end{equation}
At the minimum of this potential,
\begin{eqnarray}
{\cal A} &\sim& {\cal O}(\Lambda^{- 1/3}).
\end{eqnarray}
Thus $\Lambda^{1/6}$ is the compactification scale and is a free
parameter of the effective theory. The minimum value of ${\cal V}_{eff}$
is now the  effective four-dimensional cosmological
constant. It can be set to zero by fine-tuning the SM 3-brane
tension. (See Section 8.) 

The curvature at the minimum of the effective potential sets the mass of the
combination of $Y$-scalars determining ${\cal A}$. We can estimate
this as follows. The area ${\cal A}$  scales quadratically with the
$Y$'s, so that near the minimum of ${\cal V}_{eff}$, 
\begin{equation}
 {\cal V}_{eff} ~\sim~ \Lambda ~(\delta Y)^2.
\end{equation}
From eq. (27) we see that the canonically normalized scalar fields are
$f^2 Y$, so that eq. (34) 
 corresponds to a mass  of order $\Lambda^{1/2}/f^2$. By
 eqs. (29, ~33) and the fact that $f \sim {\cal O}(M)$, we see that this
 mass is just $1/{\cal A} M_{Pl}$. To be phenomenologically acceptable
 we must then have \cite{damour}, 
\begin{equation}
{\cal A} ~M_{Pl} < 1 {\rm cm}.
\end{equation}
This corresponds to a compactification length scale smaller than
$10^{-16}$ cm and $M$ larger than $10^{10}$ GeV. 

The mechanism considered above for acceptably stabilizing ${\cal A}$ is
minimal and attractive, but for millimeter scale compactifications it is
unacceptable and we must turn to something else.

\section{Stability from Trapped Magnetic Flux}

In this section we make use of the six-dimensional compact $U(1)$ gauge field
and six-dimensional cosmological constant to stabilize the compact
space. The mechanism is essentially a limiting case of that of
ref. \cite{cremmer} where the $U(1)$ was the remnant of a spontaneously
broken six-dimensional $SU(2)$ gauge theory. The intuitive idea is
simple.  Compact $U(1)$ gauge fields 
can have non-zero magnetic flux through closed two-dimensional
surfaces such as ${\cal M}_2$. This flux is a quantized 
topological invariant of the $U(1)$ fiber bundle structure and is therefore
fixed. This forces the flux density to increase as ${\cal
  A}$ decreases, leading to a higher magnetic energy density. This 
provides the stabilizing potential we seek. As ${\cal A}$ increases this
potential reduces, but the potential due to a small six-dimensional
cosmological constant increases, as seen in eq. (30). The size of the
compact space is determined by the balance between these two
effects. 

Explicitly, the magnetic flux through ${\cal M}_2$ is given by
\begin{equation}
\Phi \equiv \int_{{\cal M}_2} d X^4~ dX^5 
~\epsilon^{mn} {\cal F}_{mn}(X) = \frac{2 \pi N}{e},
\end{equation}
where $\epsilon^{45} = - \epsilon^{54} = 1$, $N$ must be an integer, and
$e$ is the elementary
charge that defines the abelian gauge group as a compact $U(1)$. Note
that in six-dimensions, $e$ has units of ${\rm mass}^{-1}$. Let us
briefly recall the reason for the flux quantization. Naively, the flux must
vanish by Stokes' Theorem. However this can be evaded by adding a
compensating  point-like vortex of flux, $-\Phi$,  (corresponding to the
 ``Dirac string'' of three dimensional space) somwhere on ${\cal
   M}_2$. The presence of the vortex can only be physically detected by
 the Aharnov-Bohm phase it induces in test charges, namely $- e
 \Phi$. Thus the vortex is unphysical (akin to an unphysical 
 coordinate singularity) precisely when the flux is quantized as in
 eq. (36). In this section we will consider the case $N = 1$.

The equations of motion for the gauge field following
from eq. (5), with the six-dimensional metric given by $G_{\mu \nu} =
\overline{g}_{\mu \nu}(x),~ G_{\mu m} = 0,~ G_{mn} = {\cal G}_{mn}(X^4,
X^5)$, are, 
\begin{eqnarray}
\partial_{\mu} [\sqrt{-\overline{g}}~ \overline{g}^{\mu \rho}~
\overline{g}^{\nu \sigma}~ {\cal F}_{\rho \sigma}] &=& 0, \\
\partial_{\mu} [\sqrt{-\overline{g}}~ \overline{g}^{\mu \rho} 
~{\cal F}_{\rho l}) &=& 0, \\
\partial_{m} [\sqrt{\cal G} ~{\cal G}^{mk}~ {\cal F}_{k \nu}] &=& 0, \\
\partial_m [\sqrt{\cal G}~ {\cal G}^{mk} ~{\cal G}^{nl}~ {\cal F}_{kl}]
&=& 0.
\end{eqnarray}
We will seek a solution where the gauge field has vanishing
$\mu$-components and the $m$-components are $x$-independent. Therefore
eqs. (37 -- 39) are automatically satisfied, leaving only eq. (40).

Since ${\cal F}_{mn}$ is an antisymmetric tensor it can conveniently be
written in terms of a scalar field, $B$,  on ${\cal M}_2$, 
\begin{equation}
{\cal F}_{mn} \equiv \frac{\epsilon_{mn}}{\sqrt{\cal G}} ~B,
\end{equation}
where the indices of the $\epsilon$-tensor have been lowered with the
${\cal G}_{mn}$ metric. Eq. (40) can then simplifies to, 
\begin{equation}
\partial_m B = 0.
\end{equation}
Taking into account eq. (36) (for $N = 1$) gives the solution, 
\begin{equation}
B = \frac{2 \pi}{e {\cal A}}.
\end{equation}

Substituting eqs. (41) and (43) into eq. (5) we can read off the
contribution that the magnetic energy of the trapped flux makes
 to the effective potential of the effective theory below the
 compactification scale. The result is, 
\begin{eqnarray}
{\cal V}_{eff} &=& \sum_j f^4_j ~-~ 16 \pi M^4 + \int d X^4 dX^5
\sqrt{\cal G}
[\Lambda ~+~ \frac{1}{2} B^2] \nonumber \\
&=& \Lambda ~{\cal A}(Y) ~+~ \frac{2 \pi^2}{e^2 {\cal A}(Y)} ~+~
\sum_j f^4_j ~-~ 16 \pi M^4, 
\end{eqnarray}
where we have omitted the quantum corrections discussed in the last
section since they are negligible for large compactifications.
Minimizing this effective potential for ${\cal A}$ we find that 
\begin{eqnarray}
{\cal A} &=& \frac{\sqrt{2} \pi}{e \sqrt{\Lambda}}.
\end{eqnarray}
The effective potential at this minimum will correspond to an effective
four-dimensional cosmological constant which can be made to vanish by
tuning the SM 3-brane tension. As in the previous section we can
estimate the mass of the canonically normalized combination of
$Y$-scalars corresponding to ${\cal A}(Y)$ fluctuations, and find that it
is still of order $\Lambda^{1/2}/f^2$. What is new is that we have the
extra parameter $e$ to play with, so that we can consistently arrange
for both the Compton wavelength of these fluctuations 
and the compactification length scale to be
(roughly) of order a millimeter. This phenomenologically interesting but
safe choice is accomplished by taking,
\begin{eqnarray}
e &\sim& {\cal O}(\frac{1}{\sqrt{\cal A} M^2}) ~\sim~ 
{\cal O}(\frac{1}{M_{Pl}}), \nonumber \\
\Lambda &\sim& {\cal O}(\frac{M^4}{\cal A}).
\end{eqnarray}

In this section we have treated the six-dimensional
cosmological constant and magnetic energy as perturbations to the basic
picture developed in Sections 5 and 6. Strictly, this analysis is valid
in the regime where the ${\cal M}_2$ curvature due to these new sources
is small compared to the zero-th order sources, namely the 3-branes
themselves. A six-dimensional cosmological constant is a source 
for a constant curvature of order $\Lambda/M^4$, which integrates to a
total of order ${\cal A} \Lambda/M^4$. This should be smaller
than the 3-brane deficit angles, which were order one, but not
necessarily  much 
(parametrically) smaller. Thus at the order of magnitude level this is
consistent with eq. (46). The magnetic energy balances the cosmological
constant at the minimum of the effective potential and is therefore also
consistent with the approximation we have made. In fact an exact
solution of the classical equations of motion including the magnetic
energy is possible, and agrees with what we have found.

\section{Which Cosmological Constant Problem?}

We now consider the nature of the Cosmological Constant
Problem in the present model. For a general review see
\cite{weinberg}. In fact there are
two cosmological constants that we should consider, the effective
four-dimensional constant below the compactification scale  and 
the six-dimensional constant, $\Lambda$. They pose quite different problems.

Let us begin with the four-dimensional effective cosmological constant.
It is given by the value of the effective potential at its minimum, 
\begin{equation}
{\cal V}_{eff}^{min} = 2 \Lambda {\cal A}^{min} ~+~ \sum f^4_j ~-~ 16
\pi M^4.
\end{equation}
Eq. (4) shows that, in the absence of exact supersymmetry, the
 SM vacuum energy will renormalize the corresponding 3-brane
tension by an amount roughly set by the weak scale $v$, 
\begin{equation}
f^4 = f_0^4 ~+~ {\cal  O}(v^4).
\end{equation}
Consequently, the {\it natural} size of $|{\cal V}_{eff}^{min}|$ 
is at least ${\cal O}(v^4)$. The
extreme fine tuning of $f_0$ in order to get $|{\cal V}_{eff}^{min}|$ 
to be less than the
experimental bound of $10^{-56}~v^4$, is precisely the 
usual cosmological constant problem. 
Note that in 
the absence of the  stabilization mechanism, the requirement that ${\cal
  V}_{eff}$ vanish reduces to the sum rule for the 3-brane deficit
angles, eq. (21). 

It is amusing to note that the above fine-tuning problem disappears in the case
where the extra dimensions are {\it not} compactified, as in Section 4. 
There we found a solution to the effective classical equations of
motion, where the induced metric on the 3-brane seen by the SM particles
is exactly four-dimensional Minkowksi space, without any need to fine
tune the 3-brane tension! A change in this tension only led to a change
in the deficit angle of the conical singularity in the extra
dimensions. Of course this is merely trading one major problem for
another since without compactifying ${\cal M}_2$,  gravity  remains
six-dimensional at all distances, in obvious conflict with
experiment. (For example, Newton's $1/r^2$ law is replaced by a
$1/r^4$ law.)

Let us now turn to the six-dimensional constant, $\Lambda$. The crucial
observation is that {\it the SM vacuum energy does not renormalize
  $\Lambda$}. This is because the SM fields are confined to a 3-brane
whereas $\Lambda$ represents a gravitational interaction throughout the
six-dimensional bulk spacetime. It can only be renormalized by quantum
loops of six-dimensional fields. In fact using dimensional
regularization there is no renormalization of $\Lambda$ by quantum loops of
six-dimensional gravitons and gauge fields, since they do not have a mass
scale that can appear in divergent cosmological constant terms. 
So any choice of $\Lambda$
seems technically natural. This argument requires qualification
however. Six-dimensional general relativity breaks down as an effective
field theory at scale $M$ and therefore there must be new physics by
this scale which replaces it. Therefore naturalness would require this
dynamical scale to set the size of $\Lambda$, which is much larger than we can
tolerate (see eq. (46)).
We have two choices in our effective field theory, simply accept that 
$\Lambda$ is also fine-tuned to be as small as in eq. (46), or 
consider the bulk dynamics to be supersymmetric so that a
small $\Lambda$ is technically natural. In the
latter case, we should take the view that {\it all}
 of the fundamental dynamics are
exactly supersymmetric, but that the SM sector appears non-supersymmetric
because of spontaneous supersymmetry breaking dynamics on (or by) the SM
3-brane. This supersymmetry breaking only feeds into the gravitational
sector below the compactification mass scale. 

To summarize, there are two potential cosmological constant problems,
associated with the two cosmological constants in six and four
dimensions. While the six-dimensional cosmological constant can be be
kept naturally small if the bulk dynamics is supersymmetric, this is
not an option for the effective four-dimensional cosmological constant
because we know experimentally that supersymmetry is too badly broken in
at least the SM sector. Ideas along the lines put forth in ref. 
\cite{cosmo} may be required to resolve this tough naturalness problem.

\section{Discussion}

In this paper, a 3-brane effective field theory has been constructed
which is consistent with all experimental Standard Model and
gravitational tests. The size of the compactified extra dimensions is
effectively a free parameter of the model. If it is almost a millimeter,
as proposed in ref. \cite{stanford1}, upcoming tests of short-distance gravity 
\cite{price}
will see the transition to six-dimensional gravity, while future
particle  accelerators will be sensitive to the physics of strongly
coupled gravity at the six-dimensional Planck scale, whether this is
provided by strings or something else. The present model should also have
interesting cosmological implications although we have not pursued these
here.

In the present paper, the compactification scale is effectively put in
by hand among the parameters of our starting effective theory.
Although it is
technically natural to have the compactification mass scale be much
smaller than the weak scale and six-dimensional Planck scale, thereby
realizing the proposal of ref. \cite{stanford1} for solving the Gauge
Hierarchy Problem, it would be more attractive if there were a single
dynamical scale, roughly of order the six-dimensional Planck scale $M$, 
with the
compactification scale emerging dynamically in terms of this scale
and some dimensionless parameters, $g$, say in the form $e^{-1/g^2}
M$. This would be a true elimination of the Hiererchy Problem.
It is worth exploring if a model of this type can be constructed.

\section*{Acknowledgments}
I am grateful to Jonathan Bagger for encouraging me to write up this
paper, to Andy Cohen and Martin Schmaltz for discussions, and to Nick
Evans for putting me onto ref. \cite{damour}.
This research was supported by the U.S. Department of Energy under grant
\#DE-FG02-94ER40818. 
 17

\end{document}